\documentclass[aps,prl,reprint,amsmath,amssymb,floatfix,citeautoscript]{revtex4-1}
\usepackage{graphicx}
\usepackage{amsthm}
\usepackage{bm}
\usepackage{braket} 
\usepackage{verbatim}
\usepackage{txfonts}
\usepackage{color}
\usepackage[usenames,dvipsnames]{xcolor}
\definecolor{myblue}{rgb}{0,0,1}
\usepackage[breaklinks=true,colorlinks=true,linkcolor=myblue,urlcolor=myblue,citecolor=myblue]{hyperref}

\newcommand{\vk}{{\bm{k}}}
\newcommand{\vq}{{\bm{q}}}

\begin{document}

\title{Many-Body Simulation of Two-Dimensional Electronic Spectroscopy of Excitons and Trions in Monolayer Transition-Metal Dichalcogenides}

\author{Roel Tempelaar}
\email{r.tempelaar@gmail.com}
\affiliation{Department of Chemistry, Columbia University, 3000 Broadway, New York, New York 10027, USA}

\author{Timothy C. Berkelbach}
\email{berkelbach@uchicago.edu}
\affiliation{Department of Chemistry and James Franck Institute, University of Chicago, Chicago, Illinois 60637, USA}

\begin{abstract}
We present a many-body formalism for the simulation of time-resolved nonlinear
spectroscopy and apply it to study the coherent interaction between
excitons and trions in doped transition-metal dichalcogenides.  Although the
formalism can be straightforwardly applied in a first-principles manner, for
simplicity we use a parameterized band structure and a static model dielectric
function, both of which can be obtained from a calculation using the $GW$
approximation.  Our simulation results shed light on the interplay between
singlet and triplet trions in molybdenum- and tungsten-based compounds.
Our two-dimensional electronic spectra are in excellent agreement with
recent experiments and we accurately reproduce the beating of a cross-peak
signal indicative of quantum coherence between excitons and trions.
Although we confirm that the quantum beats in molybdenum-based monolayers
unambigously reflect the exciton-trion coherence time, they are shown here to
provide a lower-bound to the coherence time of tungsten analogues due to a destructive
interference emerging from coexisting singlet and triplet trions.
\end{abstract}

\maketitle

Atomically-thin materials exhibit unique physical phenomena emerging from
extreme dimensional constraints, which adds to their attractiveness as practical
components for ultrathin electronics and optoelectronics~\cite{Novoselov_05a}.
Of particular recent interest are monolayer transition-metal dichalcogenides
(TMDCs), which are direct-bandgap analogues of graphene~\cite{Mak_10a,
Splendiani_10a}. Charge carriers in TMDCs have a large effective mass and
experience reduced dielectric screening, resulting in strong Coulomb
interactions and large exciton binding energies~\cite{Chernikov_14a, Zhu_15a, Berkelbach_18a}. 
The strong Coulomb interactions also lead to the formation of higher-order
excitonic complexes such as trions~\cite{Mak_13a, Ross_13a}, biexcitons,
\cite{You_15a}, and potentially Fermi polarons at large doping
\cite{Efimkin_17a}. Excitons are known to follow a non-hydrogenic Rydberg series
\cite{Chernikov_14a} and form in momentum valleys centered at the $K$ and $K'$
points of the Brillouin zone
with wavefunctions primarily composed of transition-metal 
$d$ orbitals~\cite{Xiao_12a}.  Such states exhibit robust valley and spin
coherence due to the sizable spin-orbit
coupling~\cite{Xiao_12a, Liu_13a}. In addition, inversion symmetry breaking
results in valley-dependent optical selection rules. In particular, circularly
polarized light has been shown to allow for valley-selective
excitation~\cite{Xiao_12a, Zheng_12a, Mak_12a}.

While the steady-state properties of TMDCs have been studied in detail
by linear optical techniques, the recent application of time-resolved
nonlinear spectroscopy has enabled the study of excited-state 
dynamics on femtosecond timescales. 
In particular, two-dimensional electronic spectroscopy (2DES)
\cite{Hybl_98a}, which has found extensive use in the study of molecular
assemblies \cite{Brixner_05a, Engel_07a, Halpin_14a, Thyrhaug_18a},
has been applied to TMDCs only quite recently \cite{Dey_15a, Moody_15a, Dey_16a,
Hao_16a, Hao_16b, Hao_17a, Hao_17b}. 
2DES is a four-wave mixing technique that improves over two-pulse pump-probe spectroscopy in its ability to map out the full third-order optical susceptibility of a sample by correlating excitation and detection frequencies. Through this approach, strong coherent interaction between excitons and trions in TMDCs have been detected, including a cross-peak beating on the 100~fs timescale~\cite{Hao_16a}.

At present time, there is a lack of first-principles techniques capable of
simulating the nonlinear optical response of condensed-phase materials. This is
in stark contrast to the state of affairs for linear spectroscopy,
where time-dependent density functional theory and the Bethe-Salpeter
equation both predict accurate spectra, including excitonic effects \cite{Albrecht_98a, Rohlfing_98a, Reining_02a, Marini_03a}.
Simulating the time-resolved nonlinear spectroscopy of trions in atomically-thin materials presents further
challenges due to the larger trion Hilbert space and the dense Brillouin
zone sampling required to resolve the dielectric function~\cite{Huser_13a, Qiu_16a}.
Here, we present a many-body computational framework for the simulation
of 2DES and apply it to the coherent interaction of trions and excitons
in monolayer TMDCs.  Although the approach can be straightforwardly implemented
in a fully first-principles manner, here we use a parameterization of the
low-energy band structure and a model dielectric function, both of which could
be obtained from a calculation using the $GW$ approximation~\cite{Hybertsen_86a}.  The
present work builds on an extension of the Bethe-Salpeter
equation to simulate linear spectra of three-body excitonic
complexes~\cite{Zhang_14a}, combined with a Brillouin zone
truncation scheme previously applied to excitons~\cite{Qiu_16a}.

\begin{figure}
\includegraphics[width=\columnwidth]{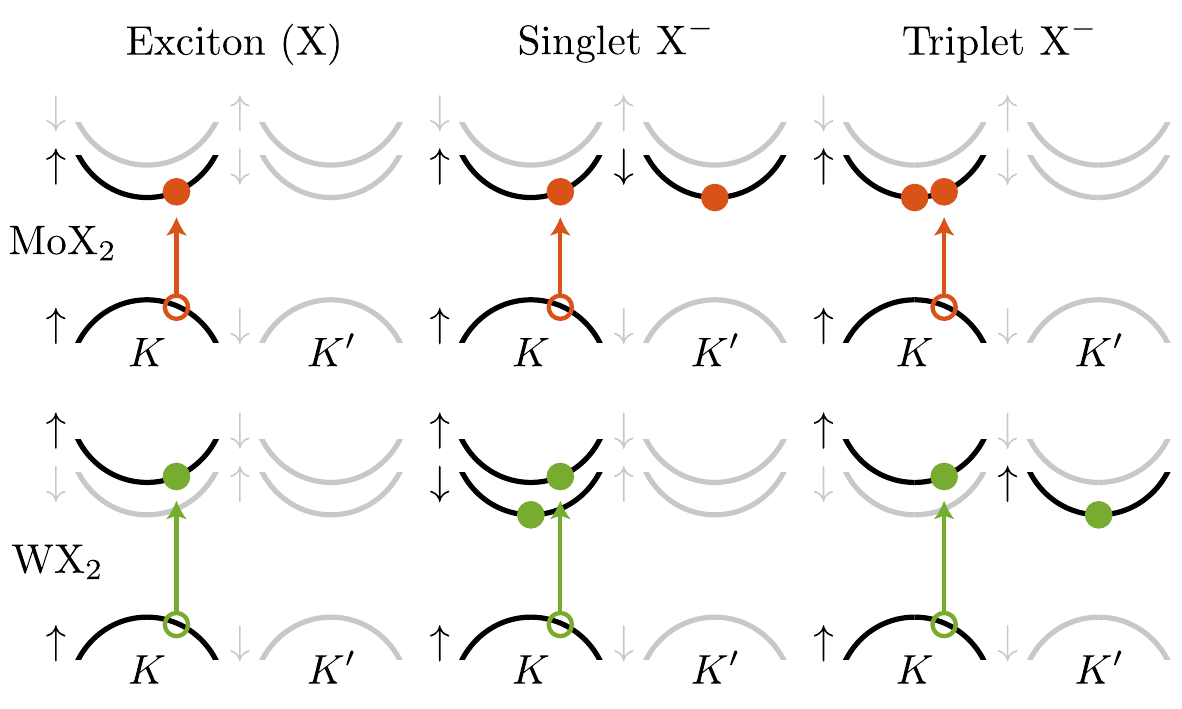}
\caption{Schematic representation of the spin dependent band structure of
monolayer MoX$_2$ and WX$_2$ near the $K$ and $K'$ points, including the
(helicity-selective) optical transitions and electronic configurations involved
in $A$ excitons and associated negatively charged trions (the lower spin-split
valence band is not shown). Singlet trions are intervalley in MoX$_2$ but
intravalley in WX$_2$; the situation is reversed for triplet trions.}
\label{Fig_Scheme}
\end{figure}

Shown in Fig.~\ref{Fig_Scheme} is a schematic of the quasiparticle band
structure of TMDCs based on molybdenum (MoX$_2$) and tungsten (WX$_2$) near the
$K$ and $K'$ points, highlighting the spin-orbit splitting of the conduction
bands \cite{Liu_13a}. The splitting of the valence bands is
an order of magnitude larger than that of the conduction bands, and results in
two distinct absorptive transitions observable in the exciton spectrum. We
restrict ourselves to the lowest-energy transition (referred to as $A$ exciton)
and its associated negatively charged trion complexes to focus on the
interpretation of experiments that energy-selectively excite this transition,
although noting that a generalization to the other transition ($B$ exciton) is
straightforward. We furthermore consider helicity selective
excitation in the $K$ valley; 
identical results would be obtained for the $K'$ valley upon flipping the spins
of the involved quasiparticles. As is well known, the combination of energy and
helicity selectivity to probe $A$ excitons in the $K$ valley effectively
corresponds to spin selective excitation, since the involved quasiparticles are
constrained to a well defined spin (spin-up, following the convention of
Fig.~\ref{Fig_Scheme}).

Negatively charged trions in TMDCs are commonly characterized based on
the spin of the two conduction-band electrons, leading to singlet or triplet
trions when the spins are opposite or equal, respectively. 
The combination of valley and spin selectivity implies selection rules 
for trions at low temperatures (with the thermal energy smaller than
the conduction band splitting). The behavior is illustrated in
Fig.~\ref{Fig_Scheme}, which considers selective excitation of the $A$
transition in the $K$ valley, creating an electron-hole pair in addition to an
initial one-electron state. At low temperatures, the initial electron is relaxed
in the minimum of either the spin-down state in the $K'$ valley or the spin-up
state in the $K$ valley. As a result, the corresponding \emph{intervalley}
trions have singlet spin whereas \emph{intravalley} trions have triplet spin.
It is easily verified that the opposite relation holds for WX$_2$.
We note that the excited electrons have identical valley and spin states only
for the triplet trion in MoX$_2$, and repulsive interactions between conduction
band electrons are therefore expected to be strongest for this case.

Because the many-body Hamiltonian conserves momentum, the trion states can be expressed as
\begin{align}
\ket{\Psi^\alpha} =\sum_{c_1,c_2,v}\sum_{\vk_1,\vk_2}C^\alpha_{c_1,c_2,v}(\vk_1,\vk_2)\;a_{v,\vk_1+\vk_2-\bm{Q}}\;a_{c_2,\vk_2}^\dagger a_{c_1,\vk_1}^\dagger\ket{0},
\end{align}
where $c_1$, $c_2$, and $v$ index the conduction and valence bands (including spin) and with $\bm{Q}$ as the momentum of the inital conduction band electron. In our simulation, details of which can be found in the Supporting Information, the band structure is described by a parameterized two-band model~\cite{Xiao_12a, Zhang_14a,Berkelbach_15a}. The trion states are calculated by configuration interaction using a many-body Hamiltonian containing a screened Coulomb interaction, as done in previous extensions of the Bethe-Salpeter equation to three-particle complexes~\cite{Zhang_14a, Deilmann_16a, Druppel_17a}. The screened Coulomb interaction is approximated as orbital-independent and isotropic, using a model dielectric function, $W(\vq) = 2\pi e^2/q\varepsilon(q)$ with $\varepsilon(q) = 1 + 2\pi \chi_\text{2D} q$, where $\chi_\text{2D}$ is a two-dimensional material-dependent polarizability~\cite{Cudazzo_11a,Berkelbach_13a, Chernikov_14a, Berkelbach_15a}. For two-dimensional materials, a very dense sampling of the Brillouin zone is required for convergence~\cite{Qiu_16a}; however, such a dense sampling makes the trion Hilbert space prohibitively large. To overcome this obstacle, we used a uniform $N\times N$ Monkhorst-Pack mesh with a cut-off radius around the $K$ and $K'$ points, denoted $k_0$. Employed previously by Qiu \emph{et al.}~for excitons~\cite{Qiu_16a}, this truncation scheme utilizes the valley confinement of low-energy excited states, and results in two convergence parameters, $N$ and $k_0$.

We first consider the exciton and trion binding energies predicted by this
approach.
Results for MoS$_2$ and WS$_2$, presented in the SM, show
the exciton binding energies to rapidly converge with $k_0$, with
near-convergence reached already for $k_0=0.10$ (in units of the inverse lattice
constant, $2\pi/a$ \footnote{The inverse lattice constants for MoS$_2$ and
WS$_2$ are 1.97 \AA$^{-1}$ and 1.96 \AA$^{-1}$, respectively.}). However, convergence with $N$ is very slow, with $N$ ranging from a
few tens to a few hundred.
Extrapolation of our results to $N=\infty$ (see SM)
yields exciton binding energies of 0.53~eV and 0.50~eV for MoS$_2$ and WS$_2$,
respectively, in fair agreement with 0.55~eV and 0.52~eV obtained in a
numerically exact diffusion Monte Carlo study of the closely-related real-space
exciton problem~\cite{Mayers_15a}. 
In contrast, the trion binding energies depend only weakly on $N$, suggesting a
cancellation of sampling errors between the total exciton and trion energies,
while a modest dependence on $k_0$ is found. For MoS$_2$, our model predicts a
singlet trion binding energy of 31~meV, whereas the triplet trion is found to be
unbound as a result of the repulsive interactions between conduction band
electrons.
These interactions are negligible for WS$_2$, where we find 
bound singlet and triplet trions with nearly identical binding
energies of 40~meV, although modest energetic splitting between these states \cite{Yu_14a, Jones_15a, Plechinger_16a, Courtade_17a} is in principle possible due to exchange interactions involving conduction and valence band electrons not considered in our model.
 The agreement with diffusion Monte Carlo results (34~meV for
singlet trions both for MoS$_2$ and W$_2$) \cite{Mayers_15a} is again
reasonable. 

\begin{figure}
\includegraphics{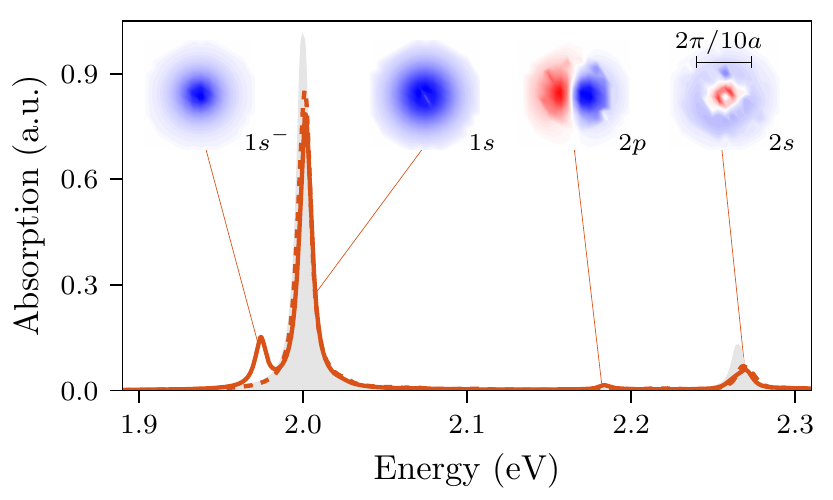}
\caption{Calculated helicity-selective singlet (solid curve) and triplet (dashed
curve) trion spectra of monolayer MoS$_2$. The two-particle exciton spectrum is
shown in grey. For selected trion states, the wavefunction in reciprocal space near the $K$ point is shown as a heatmap.}
\label{Fig_Characters}
\end{figure}

In Fig.~\ref{Fig_Characters}, we show the
zero-temperature exciton and trion linear absorption spectra for MoS$_2$,
evaluated via 
\begin{align}
S(\omega)=\frac{2\pi}{\hbar}\sum_\alpha\vert\bra{\Psi^\text{i}}V\ket{\Psi^\alpha}\vert^2\Gamma(E^\alpha-E^\text{i}-\hbar\omega).
\label{Eq_GoldenRule}
\end{align}
Here, $V=(eA/mc)\bm\lambda\cdot\bm p$ is the light-matter interaction, where $A$
is the vector potential, $e$ and $m$ are the electron charge and mass, $c$ is
the speed of light, $\bm p$ is the momentum operator, and $\bm\lambda$ is the
polarization of the optical field. Here we use circular polarization that
selectively excites carries in the $K$ valley. The exciton and trion excited
states are indexed by $\alpha$, and the associated initial state is labeled as ``i'';
while for excitons this is the vacuum state, for singlet trions the initial state has an excess spin-down electron with momentum $\bm{Q}$ at the $K^\prime$ point, and for triplet trions the initial state has an excess
spin-up electron at the $K$ point (see Fig.~\ref{Fig_Scheme}).
We note that these two initial states are isoenergetic, and so the
\textit{total} triplet spectrum would be the sum of the two associated contributions.
The function $\Gamma$, containing the excited state lifetime and other lineshape
broadening effects, is taken to be a Lorentzian with a width of 4~meV. Shown in
Fig.~\ref{Fig_Characters} are results for $N=80$ and $k_0=0.10$, while spectra
resulting from different convergence parameters are shown in the SM. 
We note that the simplified electronic structure used here (two bands
and a static, model dielectric function) combined with the Brillouin
zone truncation scheme enables us to study convergence beyond that achievable
by a fully first-principles approach and uniform
sampling~\cite{Druppel_17a}.

In Fig.~\ref{Fig_Characters}, we also show the wavefunctions in $k$-space (with an electron and hole sharing the same $k$-vector), located in the $K$ valley.
These results confirm the $s$-type azimuthal symmetry of the bound
singlet trion located at $\sim$1.97~eV. 
In addition to the trion peak, the spectrum also has an exciton resonance
at 2.00~eV, which can roughly be understood as an unbound trion in the
three-particle basis~\cite{Esser_01a}.
Moreover, the singlet trion spectrum reproduces 
the exciton spectrum into higher-energy regions of the band, displaying
the non-hydrogenic Rydberg series exhibited by excitons in
TMDCs~\cite{Chernikov_14a}, as well as the broken degeneracy between $2s$ and
$2p$ excitons~\cite{Berkelbach_15a} (the finite oscillator strength observed for the latter is the result of limited sampling, and disappears with increasing $N$). 
The same exciton features are seen in the triplet
trion spectrum,
which only differs from its singlet analogue by the absence of a bound trion
feature.  
As shown in the SM, the higher-lying
exciton states are indiscernible with a less dense sampling
of the Brillouin zone, highlighting the effectiveness of the scheme used
here.

We next turn our attention to 2DES, through which the coherent \cite{Engel_07a}
and incoherent \cite{Brixner_05a} dynamics of excited states can be monitored.
In this technique, details of which can be found elsewhere~\cite{Mukamel_1,
Jonas_03a}, a material interacts with four ultrashort laser pulses which can be
grouped into an initial ``excitation'' pair and a subsequent ``detection'' pair,
and the resulting signal is commonly presented as an excitation-detection
correlation spectrum as a function of the time delay between the two pulse
pairs. Different combinations of pulse interactions result in different spectral
signals. In our aim to interpret recent experiments on TMDCs \cite{Hao_16a}, we
specifically focus on the non-rephasing stimulated emission signal,
\begin{align}
&S(\omega_1,t_2,\omega_3)=-\Big(\frac{2\pi}{\hbar}\Big)^2\sum_{\alpha,\beta}\vert\bra{\Psi^\text{i}}V\ket{\Psi^\alpha}\vert^2\vert\bra{\Psi^\text{i}}V\ket{\Psi^\beta}\vert^2\nonumber\\
&\times e^{-(i\omega_{\alpha\beta}+\gamma_{\alpha\beta})t_2}\Gamma^*(E^\alpha-E^\text{i}-\hbar\omega_1)\Gamma(E^\beta-E^\text{i}-\hbar\omega_3).
\label{Eq_2DES}
\end{align}
Here, $\omega_1$ and $\omega_3$ are the excitation and detection energies,
respectively, and $t_2$ is the time delay.
The excitation energy difference between excited 
states $\alpha$ and $\beta$ is given by
$\omega_{\alpha\beta}=(E_\alpha-E_\beta)/\hbar$ and a phenomelogical 
decoherence rate is given by $\gamma_{\alpha\beta}$. The resulting damped
oscillation (quantum beat) is mapped onto the 2DES signal weighted by
the product of transition matrix elements between the excited states and a 
common initial state $\Psi^\text{i}$.  In particular, excited states
that do not share a common initial state, as might arise in inhomogeneous
samples, do not show coherent cross peaks in 2DES.

\begin{figure}
\includegraphics{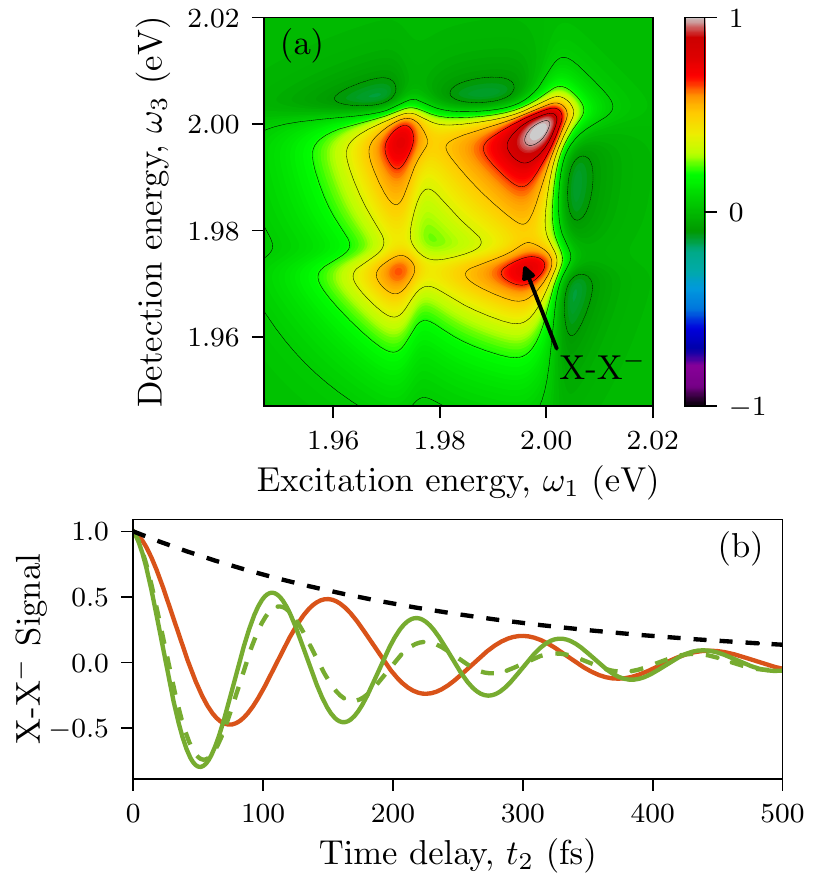}
\caption{(a) Total (singlet plus triplet) trion two-dimensional electronic
spectrum of monolayer MoS$_2$, calculated through Eq.~\ref{Eq_2DES} at zero time
delay ($t_2=0$~fs). (b) Time-dependent signal at the lower cross-peak (X-X$^-$)
location for MoS$_2$ (red) and WS$_2$ (green), shown together with an
exponential indicating the phenomenological exciton-trion coherence decay (black
curve). For WS$_2$, results are shown for fully degenerate singlet and triplet trions (solid), and for which the triplet trion is blue-shifted by 7~meV as a result of exchange interactions (dash).}
\label{Fig_2DSpectrum}
\end{figure}

Recently, Hao \emph{et al.}~recorded oscillatory signals in 2DES of
electron-doped MoSe$_2$ monolayers at 20~K \cite{Hao_16a}, yielding indications
of coherent interactions between the trion and the $1s$ exciton, and the
observed quantum beat decay suggested an associated dephasing time of
$\gamma_{\text{X-X}^-}^{-1}=250$~fs. However, a reliable determination of the exciton-trion coherence time requires detailed knowledge on how such quantum beats are affected by possible interfering oscillatory signals. Our many-body formalism, in its ability to simulate 2DES, allows to address this in a straightforward manner, while offering the prospect of microscopically investigating the decoherence mechanisms (e.g., through a Redfield theory treatment of dissipative degrees of freedom). For now we resort to a phenomenological treatment of the latter, and study the 2DES of TMDCs using MoS$_2$ and WS$_2$ as
representative examples. In Fig.~\ref{Fig_2DSpectrum}(a), we show the sum of the singlet and triplet 2D spectrum for MoS$_2$ resulting from
Eq.~\ref{Eq_2DES} at zero time delay and with cocircularly polarized pulses,
obtained for the same convergence parameters as in Fig.~\ref{Fig_Characters}.
The impulsive signal is multiplied by a Gaussian laser spectrum centered at the
bound trion state and with a standard deviation of 17~meV, accounting for the
limited laser bandwidth affecting the experimental measurements.~\cite{Hao_16a}
Apart from a rigid spectral shift through which MoS$_2$ differs from MoSe$_2$,
the agreement with the 2DES measurements by Hao \emph{et al.}~is excellent, with
the spectrum showing four peaks in a square arrangement, resulting from two
optical transitions readily identified as the bound trion and $1s$ exciton.

According to Eq.~\ref{Eq_2DES}, the quantum beats due to the 
exciton-trion coherence are mapped onto the cross-peaks corresponding to trion
excitation and exciton detection, and \emph{vice versa}. Indeed, these spectral
locations were employed in the quantum beat measurements by Hao \emph{et al.}~\cite{Hao_16a}. However, as discussed above, 
the quantum beats only result from pairs of states
($\alpha$ and $\beta$) that are optically coupled to a common initial state,
$\Psi^\text{i}$. The $1s$ exciton is observed in all of the trion and exciton
calculations, but only its resonance in the \textit{singlet} trion configuration
contributes to exciton-trion quantum beats observed for MoX$_2$, since only this
resonance optically couples to the same initial state as the bound trion
(a spin-down electron relaxed in the $K'$ valley). 

Fig.~\ref{Fig_2DSpectrum}(b) shows the time-dependent signal of the lower
(below-diagonal) cross-peak (LC) for MoS$_2$ resulting from our model while
imposing
$\gamma_{\text{X-X}^-}^{-1}=250$~fs (the other cross-peak yields an identical
signal except for incoherent contributions from population transfer that are not
considered in our simulation). The signal features a pronounced oscillation,
consistent with the measurements, with the oscillation period matching the
Fourier inverse of the singlet trion binding energy,
$T = h/(31~\mathrm{meV}) = 133$~fs. Consistent with the above
discussion, we find this quantum beat to result from the bound singlet trion state
coherently interacting with its exciton resonance. A comparison of the
associated quantum beat decay with the reference decay function
$e^{-t_2\gamma_{\text{X-X}^-}}$ shows the destructive interference with auxiliary
states to be negligible, such that the exciton-trion coherence time is
indeed accurately reflected in this oscillatory signal. This substantiates that
quantum dephasing in MoSe$_2$ induces a measurable coherence decay time of
250~fs, as inferred from the reported 2DES experiments \cite{Hao_16a}.

For WX$_2$, both singlet and triplet trions form bound states, and as such both
contribute to exciton-trion quantum beats resulting from coherent interactions
with their respective exciton resonance. 
In our simulation, we can choose to include or exclude the repulsive intravalley
exchange interaction, which breaks the degeneracy of the singlet and triplet
trions.
If we exclude the interaction, the beating pattern shown in
Fig.~\ref{Fig_2DSpectrum}(b) is very similar to the MoS$_2$ quantum beat, apart
from a slightly higher oscillation frequency (consistent with the slightly
higher binding energy). 
Again, the quantum beat decay is found to form a reliable probe of
the underlying exciton-trion dephasing time. 
However, if we include a $k$-independent intravalley exchange interaction of
7~meV \cite{Yu_14a, Jones_15a, Plechinger_16a, Courtade_17a}, which leads to an energy splitting of 7~meV, the beating signal
changes appreciably, as shown in Fig.~\ref{Fig_2DSpectrum}(b).
The nondegenerate trion states lead to an apparent destructive interference in
the total (singlet plus triplet) quantum beat, as a result of which the beat
decay occurs considerably faster than the actual dephasing time. Taken together,
these results demonstrate that the quantum beats observed at the exciton-trion
cross-peak locations in 2DES provide a lower bound on the actual exciton-trion
coherence time, and that WX$_2$ in particular warrants caution because of the
presence of two (nearly degenerate) trion species.

In conclusion, we have presented a many-body formalism for the simulation
of time-resolved nonlinear spectroscopy including three-particle
excitonic complexes.  Although the formalism can be straightforwardly
implemented in a first-principles manner, we have here employed a parameterized
two-band model and an isotropic, static dielectric function.  Combined with
a careful truncation of the Brillouin zone, these choices allowed us
to provide highly converged results despite the otherwise high computational
cost.
In applying this
formalism to excitons and trions, we uncover various fundamental properties of
these charge carrier complexes that relate to the optoelectronic functionality
of TMDCs and provide excellent agreement with recently measured 2DES.
As noted before, helicity and frequency selective excitation of the $A$
transition in the $K$ valley allows control of the spin state of the optically
created electron-hole pair. Consistent with our results, an even more
comprehensive spin control can be achieved for bound trion states in MoX$_2$:
helicity selective excitation at the bound trion transition generates three-body
complexes consisting of a spin-up hole, and spin-differing electrons (following
the convention from Fig.~\ref{Fig_Scheme}). The resulting state coherently
interacts with exciton resonances that optically couple to a shared ground state
consisting of a spin down electron relaxed in the $K'$ valley. In case of
WX$_2$, where both singlet and triplet trions form bound states with
near-degenerate transition energies, such selective excitation generates both
well-defined spin configurations with ratios dictated by the spins of the doping
charges, and each triplet state coherently interacts with the exciton resonance
with which it shares a one-electron ground state. In real space, such a sharing
of a ground state can be thought of as the photoexcited electron-hole pair and
the single electron residing within each others coherent domain. This is
automatically fulfilled in theoretical models based on Bloch states,
representative of pristine, extended crystals with translational symmetry, such
as employed here. Nevertheless, actual materials are characterized by a certain
degree of impurities and scattering with phonons, which break this symmetry and
limits the size of coherent domains. 
Especially at the level of theory employed here, extension of our framework
to include electron-phonon coupling is straightforward and would allow to
unravel the microscopic origin of electronic decoherence and relaxation. 

This work was supported in part by the Air Force Office of Scientific Research under award number FA9550-18-1-0058. T.C.B.~is an Alfred P.~Sloan Research Fellow. We thank Alexey Chernikov and David Reichman for helpful discussions.

\bibliography{Bibliography}

\end{document}